\documentclass[lettersize,journal]{IEEEtran}
\usepackage{amsmath,amsfonts}
\usepackage{algorithmic}
\usepackage{algorithm}
\usepackage{array}
\usepackage[caption=false,font=normalsize,labelfont=sf,textfont=sf]{subfig}
\usepackage{textcomp}
\usepackage{stfloats}
\usepackage{url}
\usepackage{verbatim}
\usepackage{graphicx}
\usepackage{cite}

\usepackage{enumitem}
\usepackage{listings}
\usepackage{multirow}

\begin{document}

\title{Zarr-Based Chunk-Level Cumulative Sums in Reduced Dimensions}

\author{
        Hailiang Zhang, Dieu My T. Nguyen, Christine Smit, Mahabal Hegde
\thanks{All authors are with the Goddard Earth Science Data and Information Science Center, National Aeronautics and Space Administration, Greenbelt, MD, USA 20771. (email: hailiang.zhang@nasa.gov, dieumy.t.nguyen@nasa.gov, mahabaleshwa.s.hegde@nasa.gov, christine.e.smit@nasa.gov).}
\thanks{Hailiang Zhang and Dieu My T. Nguyen are with Adnet Systems, Inc. Christine Smit is with Telophase Co.}}

\markboth{}%
{Shell \MakeLowercase{\textit{et al.}}: A Sample Article Using IEEEtran.cls for IEEE Journals}

\IEEEpubid{}

\maketitle

\begin{abstract}
Data analysis on massive multi-dimensional data, such as high-resolution large-region time averaging or area averaging for geospatial data, often involves calculations over a significant number of data points. While performing calculations in scalable and flexible distributed or cloud environments is a viable option, a full scan of large data volumes still serves as a computationally intensive bottleneck, leading to significant cost. This paper introduces a generic and comprehensive method to address these computational challenges. This method generates a small, size-tunable supplementary dataset that stores the cumulative sums along specific subset dimensions on top of the raw data. This minor addition unlocks rapid and cheap high-resolution large-region data analysis, making calculations over large numbers of data points feasible with small instances or even microservices in the cloud. This method is general-purpose, but is particularly well-suited for data stored in chunked, cloud-optimized formats and for services running in distributed or cloud environments. We present a Zarr extension proposal to integrate the specifications of this method and facilitate its straightforward implementation in general-purpose software applications. Benchmark tests demonstrate that this method, implemented in Amazon Web services (AWS), significantly outperforms the brute-force approach used in  on-premises services. With just 5\% supplemental storage, this method achieves a performance that is 3-4 orders of magnitude ($\sim$10,000 times) faster than the brute-force approach, while incurring significantly reduced computational costs.
\end{abstract}


\section{Introduction}


\IEEEPARstart{P}{erforming} calculations over billions of data points is computationally expensive. Cloud computing offers a viable solution, providing accessible, cost-effective, scalable, and flexible storage and computational resources, effectively alleviating the burden of managing infrastructure for resource-intensive tasks. Nevertheless, even with parallelized operations, a comprehensive scan over all data points is typically necessary, serving as an IO-intensive and computationally expensive bottleneck within any computational infrastructure. In dealing with the broad scope of performing diverse calculations across vast datasets, caching—temporarily storing frequently accessed data—can help reduce the computational load, but the substantial storage overhead remains a significant concern.

While processing massive datasets presents significant challenges, certain types of algorithms offer opportunities for optimization. This work is motivated by Giovanni (Geospatial Interactive Online Visualization and Analysis Infrastructure) at NASA’s GES DISC, a web-based platform for visualizing, analyzing, and intercomparing multisensor Earth science data \cite{berrick_2008}. Historically running on a single on-premise server, Giovanni is now transitioning to Amazon Web Services (AWS) for serverless data storage and computing infrastructure, enabling cost savings and improved performance through scalable and elastic resources \cite{hegde_2017, ramachandran_2017}.

Giovanni handles a large volume of daily user requests for analysis and visualization, with over 80\% focused on time-averaged maps and area-averaged time series. Both services involve averaging data over contiguous time ranges (for maps) or spatial regions (for time series), requiring a comprehensive data scan. Calculating the cumulative sum is the primary effort for averaging across one or more dimensions. If we pre-calculate these cumulative sums, the computationally demanding work is performed just once, and these pre-calculated sums can subsequently be used to make real-time on-demand calculations much more efficient. To illustrate this, consider a 1-dimensional case with the sequence $A = \{a, b, c, d\}$, where the cumulative sums are $S = \{s_0, s_1, s_2, s_3\}$, with $s_0 = a$, $s_1 = a + b$, $s_2 = a + b + c$, and $s_3 = a + b + c + d$. Using these pre-calculated sums, we can speed up the averaging by simply taking the boundary values from the cumulative sums, rather than reading all the data values within the boundary. For instance, assuming zero-based indexing, averaging over the range $[1:]$ is calculated as $A[1:] = (s_3 - s_0) / 3$ by using just 2 boundary values, instead of calculating $(b + c + d) / 3$ by reading all 3 values.

Previously, we proposed a method for Giovanni by using multi-dimensional cumulative sums to provide fast and cost-efficient analysis, including area and time averaging, on the cloud \cite{zhang_2019}. Compared to the standard brute-force method, this approach dramatically reduced computation time. For example, we reduced computational time from minutes to seconds for a 10-year area average over daily data of spatial resolution of 1x1 degree while incurring a reasonable cloud cost of only \$5 for 100,000 service requests. However, while this method greatly improved the performance of data analysis, generating the cumulative sums at each data point incurred high storage and data preparation overhead as well as higher susceptibility to data overflow.

In this paper, we improve upon the previous method and introduce a new method that allows for more efficient generation of cumulative sums, in addition to improved performance in data analysis. This method provides multi-dimensional averaging services for data stored in Zarr \cite{abernathey_2018}, a cloud-optimized chunked storage format, with little additional computation and storage overhead and without modifying the \newline \newline original raw  data or impacting the existing analysis services. Instead of accumulating all data values as commonly done, our method computes the cumulative sums at the chunk level in stepwise-reduced dimensions on the regular grid and creates a small adjustable set of auxiliary data. While the development of this method was motivated by Giovanni’s problem, the method is generic and dimension-agnostic, making it applicable to any chunked multi-dimensional data. In the following sections, we describe this method and its key features, present its results, and compare its performance against the brute-force method (i.e., a full scan of the data). As cloud-based data infrastructures become more common, this work contributes a new optimized approach alongside open-source software to address common data analysis requirements.

\section{Method}
This manuscript presents a method for fast and cost-efficient data analysis across large contiguous regions of massive multi-dimensional arrays. While the method is general-purpose, it is designed to work particularly well for chunked data in distributed or cloud environments. This method introduces small, tunable supplementary data that stores the cumulative sums along specific dimension(s) on top of the raw data. The small addition of supplementary data turns high-resolution large-region data analysis, which is usually resource-intensive and time-consuming, into analysis that can be run rapidly with small, cheap instances or even microservices in the cloud, such as AWS lambda. We demonstrate this method using Zarr, a popular cloud-optimized chunked storage format, and hence call it ``Zarr-based chunk-level cumulative sums in reduced dimensions''. For simplicity, the term ``accumulation'' is used interchangeably with ``cumulative sums'' throughout this manuscript.

The method will be presented in the following two phases: Accumulation Data Generation, which describes how to create accumulation data, and Accumulation Data Analysis, which explains how to use this data to perform fast and cost-effective data analysis.

\subsection{Accumulation Data Generation}
The first phase is to generate the accumulation data. The size of this data is significantly smaller than the raw data, often just a small fraction, because it is limited to a small subset of data points for only a selected subset of dimensions:
\begin{enumerate}
    \item Cumulative sums are only computed along a selected subset of dimensions, which can be chosen based on the application requirements. There is no need to pre-calculate cumulative sums that will never be used.
    \item Cumulative sums are only computed at a small subset of data points, contingent on the selected subset of dimensions.
\end{enumerate}

The criteria for selecting a subset of dimensions based on the application requirements will be provided in the next section (“Accumulation Data Analysis”). This section focuses on selecting a subset of data points for accumulation based on the given subset of dimensions. A 3-dimensional (3-D) example is used for illustration, although this method is applicable to data with any number of dimensions.

\begin{figure}[!t]
\centering
\includegraphics[width=0.8\linewidth]{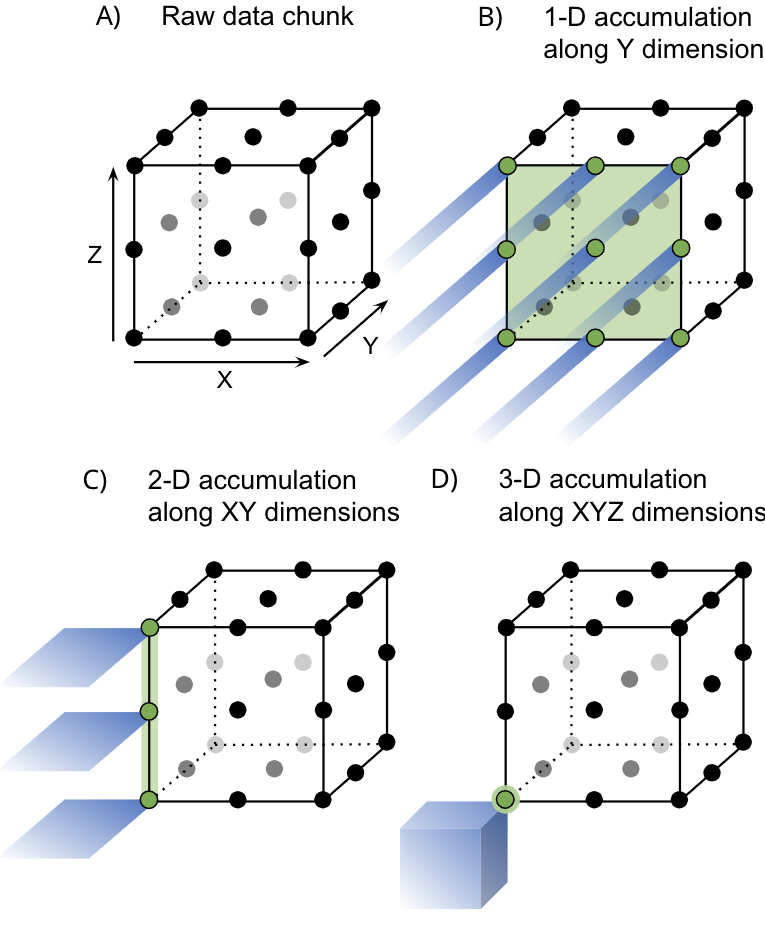}
\caption{Chunk-level Accumulation Data Generation. A) A 3-D raw data chunk along ``X'', “Y”, “Z” dimensions. For clarity, only three data points are shown along each dimension, although a typical chunk contains a significantly larger number of data points.  B) 1-D accumulation. For data points on the foremost 2-D “XZ” plane (highlighted in green), 1-D accumulation is performed along its complementary “Y” dimension (shown in blue). C)  2-D accumulation. For data points on the foremost 1-D edge along the “Z” axis (highlighted in green bar), 2-D accumulation is performed along the complementary “XY” dimensions (shown in blue). D) 3-D accumulation. For the single lower left data point (highlighted in green circle), 3-D accumulation is performed along all three dimensions (shown in blue).}
\label{fig_1}
\end{figure}

As a general rule, accumulation is performed only for data points on the foremost geometrical elements of a chunk, with cumulative sums computed along their complementary dimensions. Consider a 3-D Zarr dataset chunked along “X”, “Y” and “Z” dimensions (Figure 1-A) where accumulation is performed across the following subset dimensions:  

\begin{enumerate}[leftmargin=*]
    \item 1-D accumulation: For data points on the foremost 2-D planes (front, left, and bottom) of a chunk, accumulation is performed along their complementary dimensions, resulting in 1-D accumulation:

    \begin{align}
    \mathrm{acc}_{m}^{\mathbf{X}}[j,k] &= \sum_{i=0}^{C_{\mathbf{X}} \cdot (m+1)} \mathrm{data}[i,j,k] \notag \\
    \mathrm{acc}_{n}^{\mathbf{Y}}[i,k] &= \sum_{j=0}^{C_{\mathbf{Y}} \cdot (n+1)} \mathrm{data}[i,j,k] \notag \\
    \mathrm{acc}_{p}^{\mathbf{Z}}[i,j] &= \sum_{k=0}^{C_{\mathbf{Z}} \cdot (p+1)} \mathrm{data}[i,j,k] \label{eq:eq_1}
    \end{align}

    Here, $acc$ stands for accumulation, ($i$, $j$, $k$), ($m$, $n$, $p$), and ($C_X$, $C_Y$, $C_Z$) are respectively the data indices, chunk indices, and chunk sizes along X, Y and Z dimensions. This notation will be consistently used throughout the paper. For example, as shown in Figure 1-B, for each point on the front “XZ” plane (highlighted in green plane), cumulative sums are computed along its complementary “Y” dimension (shown in blue). \\

    \item 2-D accumulation: For data points on the foremost 1-D edges of a chunk, accumulation is performed along their complementary dimensions, resulting in 2-D accumulation:
    \begin{align}\mathrm{acc}_{m,n}^{\mathbf{X},\mathbf{Y}}[k] &= \sum_{i=0}^{C_{\mathbf{X}} \cdot (m+1)} \sum_{j=0}^{C_{\mathbf{Y}} \cdot (n+1)} \mathrm{data}[i,j,k] \notag \\
    \mathrm{acc}_{m,p}^{\mathbf{X},\mathbf{Z}}[j] &= \sum_{i=0}^{C_{\mathbf{X}} \cdot (m+1)} \sum_{k=0}^{C_{\mathbf{Z}} \cdot (p+1)} \mathrm{data}[i,j,k] \notag \\
    \mathrm{acc}_{n,p}^{\mathbf{Y},\mathbf{Z}}[i] &= \sum_{j=0}^{C_{\mathbf{Y}} \cdot (n+1)} \sum_{k=0}^{C_{\mathbf{Z}} \cdot (p+1)} \mathrm{data}[i,j,k] \label{eq:accumulation2}
    \end{align}

    For example, as shown in Figure 1-C, for each point on the foremost edge along the “Z” axis (highlighted in green bar), cumulative sums are computed along its complementary “XY” dimensions (shown in blue). \\

    \item 3-D accumulation: For the single foremost data point of a chunk, accumulation is performed along its complementary dimensions, resulting in 3-D accumulation. As shown in Figure 1-D, for the foremost data point within the chunk (highlighted in green circle), cumulative sum is computed along its complementary “XYZ” dimension (shown in blue).
    \begin{equation}\mathrm{acc}_{m,n,p}^{\mathbf{X},\mathbf{Y},\mathbf{Z}} = \sum_{i=0}^{C_{\mathbf{X}} \cdot (m+1)} \sum_{j=0}^{C_{\mathbf{Y}} \cdot (n+1)} \sum_{k=0}^{C_{\mathbf{Z}} \cdot (p+1)} \mathrm{data}[i,j,k]
    \end{equation}
\end{enumerate}

While cumulative sums may introduce storage and data preparation overhead, it is minimized in our proposed Zarr-based chunk-level accumulation method for the following reasons:

\begin{enumerate}
    \item Given that a typical Zarr chunk contains a large number of data points, the points selected for accumulation – being on the foremost geometrical elements – constitute a very small fraction of the entire dataset.
    \item As more dimensions are targeted for accumulation, fewer data points are required for cumulative sum computation. Specifically, for the intensive 3-D accumulation from the example above, only one point is needed per chunk. 
    \item Cumulative sums computed for lower dimensions can be efficiently re-utilized for higher dimensional accumulation, leading to notable reduction in accumulation data preparation time.
\end{enumerate}

Moreover, as will be discussed in later sections (“Zarr extension proposal”), accumulation can be performed over specific intervals of chunks, rather than each single chunk, allowing for customizable minimization of accumulation overhead.

\subsection{Accumulation Data Analysis}
Despite incurring a minimal and adjustable overhead in storage and data generation, this slight overhead yields substantial performance improvements and significant cost reduction for the analysis of large, multi-dimensional arrays.

\begin{figure}[!t]
\centering
\includegraphics[width=\linewidth]{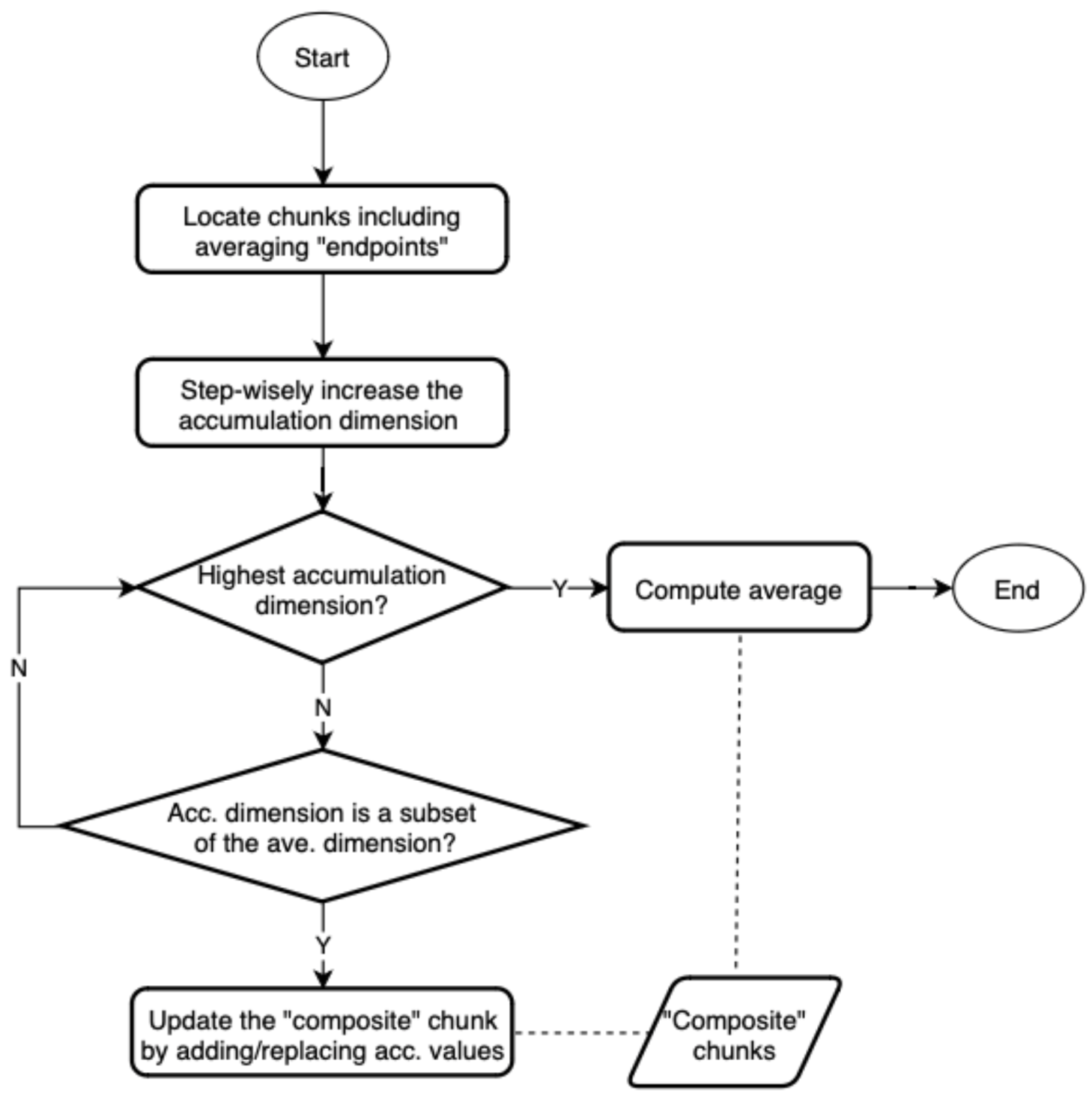}
\caption{Workflow for data analysis based on the method of “Zarr-based chunk-level cumulative sums in reduced dimensions”. “Y” indicates “yes” and “N” indicates “no”.}
\label{fig_2}
\end{figure}

The proposed accumulation-based approach introduces a generic workflow, shown in Figure 2, that is applicable to various aggregation service types and any dimension combinations. This workflow reconstructs chunks at aggregation endpoints into a “composite” chunk – a hybrid of accumulation and raw data. Data analysis is then performed through linear combinations of data within these ”composite” chunks. By limiting data extraction and computation to a minimal number of endpoint chunks along the aggregation dimension(s)—typically one to a few—this method significantly reduces network transactions and computation load, thereby enabling fast and cost-effective data analysis in distributed or cloud environments.

The workflow is demonstrated through the following two examples.

\subsubsection{Area averaging} 

First consider a global area averaging of 3-D geospatial data across “latitude” (“lat” for short), “longitude” (“lon” for short), and “time” dimensions. The averaging dimension set in this case is {“lat”, “lon”}, with averaging endpoints typically located at geospatial coordinates (90, 180). Figure 3-A shows a chunk containing these endpoints, marked by a red bar. The workflow starts with the raw data chunk, upon which the “composite” chunk is constructed by systematically integrating the accumulation data through the following sequential steps, each step incrementally increasing the accumulation dimension numbers.

\begin{enumerate}[label=\roman*.]
    \item 1-D accumulation integration. We begin with the 1-D accumulation data, corresponding to the points on the chunk's three foremost surface planes (front, left, and bottom). For each plane, if its accumulation dimension is a subset of the averaging dimension set, the corresponding raw data will be replaced with accumulation values. As illustrated in Figure 3-B, the accumulation dimensions of the front and left plane–\{“lat”\} and \{“lon”\} respectively–are subsets of the averaging dimension set \{“lat”, “lon”\}, thus the raw data on these 2 planes are replaced with 1-D accumulation values. However, the bottom plane's accumulation dimension, \{“time”\}, is not a subset of the averaging dimension set, and therefore the raw data remain unchanged. Note that the ambiguity in data assignment at the intersection of the front and left planes are resolved in subsequent steps.

    \item 2-D accumulation integration. Next we proceed to the 2-D accumulation data, corresponding to the points on the leading edges of the chunk. As illustrated in Figure 3-C, only the accumulation dimension of the edge along “time“ axis (i.e. {“lat”, “lon”}) is a subset of (and actually equal to) the averaging dimension set {“lat”, “lon”}, and therefore the data points in the “composite” chunk from the previous step are further replaced with the corresponding 2-D accumulation values. This replacement also resolves the data assignment ambiguity noted in the previous step.

    \item 3-D accumulation integration. The final step in the “composite” chunk construction is from the 3-D accumulation data.  However, since its accumulation dimension {“lat”, “lon”, “time”} is not a subset of the averaging dimension {“lat”, “lon”}, no replacements occur in the “composite” chunk in this step.
\end{enumerate}

\begin{figure}[!t]
\centering
\includegraphics[width=\linewidth]{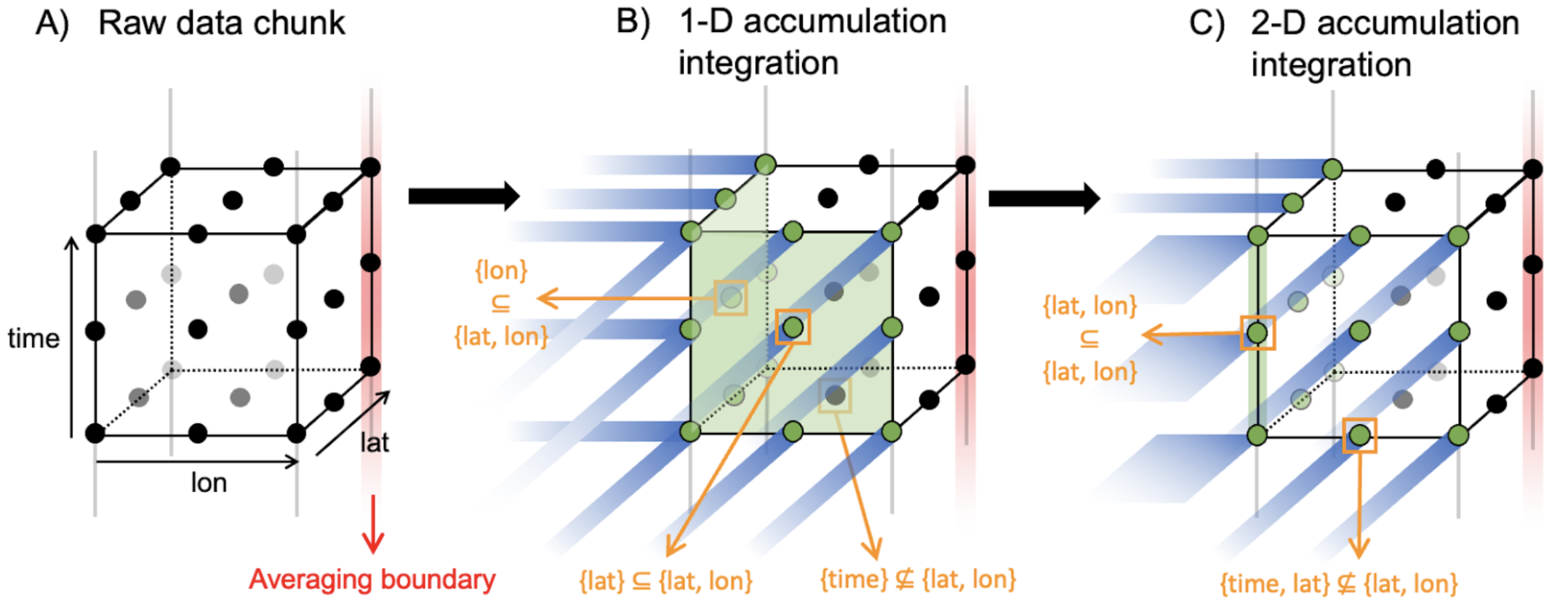}
\caption{Steps for constructing the ``composite'' chunk for accumulation-based area averaging service. A) Original 3-D data chunk with ``latitude'' (labeled ``lat'' for short), ``longitude'' (labeled ``lon'' for short), and ``time'' dimensions. The area averaging is performed along {``lat'', ``lon''} dimensions, with the averaging boundary marked by a red bar. B) 1-D accumulation integration into the ``composite'' chunk. Data points within the front and left plane (highlighted in green plane), not the bottom plane, have accumulation dimensions as a subset of the averaging dimensions (illustrated in orange), and therefore the corresponding 1-D accumulation data is integrated into the ``composite'' chunk. C) 2-D accumulation integration into the ``composite'' chunk. The accumulation dimensions for the data points on the edge along the ``time'' axis (highlighted in green bar) are subsets of the averaging dimensions (illustrated in orange), and are therefore integrated into the ``composite'' chunk. The accumulation dimension for the data points on the edge along the ``lat'' and ``lon'' axes are not subsets of the averaging dimension (illustrated in orange) and therefore the raw data remains untouched.}
\label{fig_3}
\end{figure}

The above steps complete the construction of the “composite” chunks, based on which the averaging results can be readily derived through linear combination of the data within these chunks. For global averaging time series, the averaging result for each time slice is obtained by simply summing the data values within the “composite” chunk at the endpoint – including the 1-D and 2-D accumulation data from the foremost geometrical elements as determined from the workflow, and the raw data elsewhere (Figure 4) -- and then divided by the total number of data points from the origin:
\begin{align}
\mathrm{avg}\big[i_{\boldsymbol{\mathrm{lat}}}^{\text{end}}, i_{\boldsymbol{\mathrm{lon}}}^{\text{end}}\big] 
&= \frac{\mathrm{sum}\big[i_{\boldsymbol{\mathrm{lat}}}^{\text{end}}, i_{\boldsymbol{\mathrm{lon}}}^{\text{end}}\big]}{i_{\boldsymbol{\mathrm{lat}}}^{\text{end}} \cdot i_{\boldsymbol{\mathrm{lon}}}^{\text{end}}} \nonumber \\
&= \Bigg( 
    \mathrm{acc}_{\lfloor i_{\boldsymbol{\mathrm{lat}}}^{\text{end}}/C_{\boldsymbol{\mathrm{lat}}} \rfloor-1,\ \lfloor i_{\boldsymbol{\mathrm{lon}}}^{\text{end}}/C_{\boldsymbol{\mathrm{lon}}} \rfloor-1}^{\boldsymbol{\mathrm{lat}}, \boldsymbol{\mathrm{lon}}} \nonumber \\
&\quad + \sum_{i = \lfloor i_{\boldsymbol{\mathrm{lat}}}^{\text{end}}/C_{\boldsymbol{\mathrm{lat}}} \rfloor}^{i_{\boldsymbol{\mathrm{lat}}}^{\text{end}}}
    \mathrm{acc}_{\lfloor i_{\boldsymbol{\mathrm{lon}}}^{\text{end}}/C_{\boldsymbol{\mathrm{lon}}} \rfloor-1}^{\boldsymbol{\mathrm{lon}}}[i] \nonumber \\
&\quad + \sum_{j=\lfloor i_{\boldsymbol{\mathrm{lon}}}^{\text{end}}/C_{\boldsymbol{\mathrm{lon}}} \rfloor}^{i_{\boldsymbol{\mathrm{lon}}}^{\text{end}}}
    \mathrm{acc}_{\lfloor i_{\boldsymbol{\mathrm{lat}}}^{\text{end}}/C_{\boldsymbol{\mathrm{lat}}} \rfloor-1}^{\boldsymbol{\mathrm{lat}}}[j] \nonumber \\
&\quad + \sum_{i=\lfloor i_{\boldsymbol{\mathrm{lat}}}^{\text{end}}/C_{\boldsymbol{\mathrm{lat}}} \rfloor}^{i_{\boldsymbol{\mathrm{lat}}}^{\text{end}}} 
    \sum_{j=\lfloor i_{\boldsymbol{\mathrm{lon}}}^{\text{end}}/C_{\boldsymbol{\mathrm{lon}}} \rfloor}^{i_{\boldsymbol{\mathrm{lon}}}^{\text{end}}} 
    \mathrm{data}[i,j] 
\Bigg) \nonumber \\
&\quad \times \frac{1}{i_{\boldsymbol{\mathrm{lat}}}^{\text{end}} \cdot i_{\boldsymbol{\mathrm{lon}}}^{\text{end}}}.
\end{align}

Here $(i^{end}_{lat}, i^{end}_{lon})$ are the spatial indices of the averaging endpoints, and ($C_{lat}$, $C_{lon}$) are the chunk sizes along latitude and longitude dimensions; it’s obvious that $(\lfloor i_{\boldsymbol{\mathrm{lat}}}^{\text{end}}/C_{\boldsymbol{\mathrm{lat}}} \rfloor$, $\lfloor i_{\boldsymbol{\mathrm{lon}}}^{\text{end}}/C_{\boldsymbol{\mathrm{lon}}} \rfloor)$ are the indices of the chunk where the averaging endpoint resides (Figure 4). Note that the time indices are omitted here for clarity.

\begin{figure}[!t]
\centering
\includegraphics[width=0.7\linewidth]{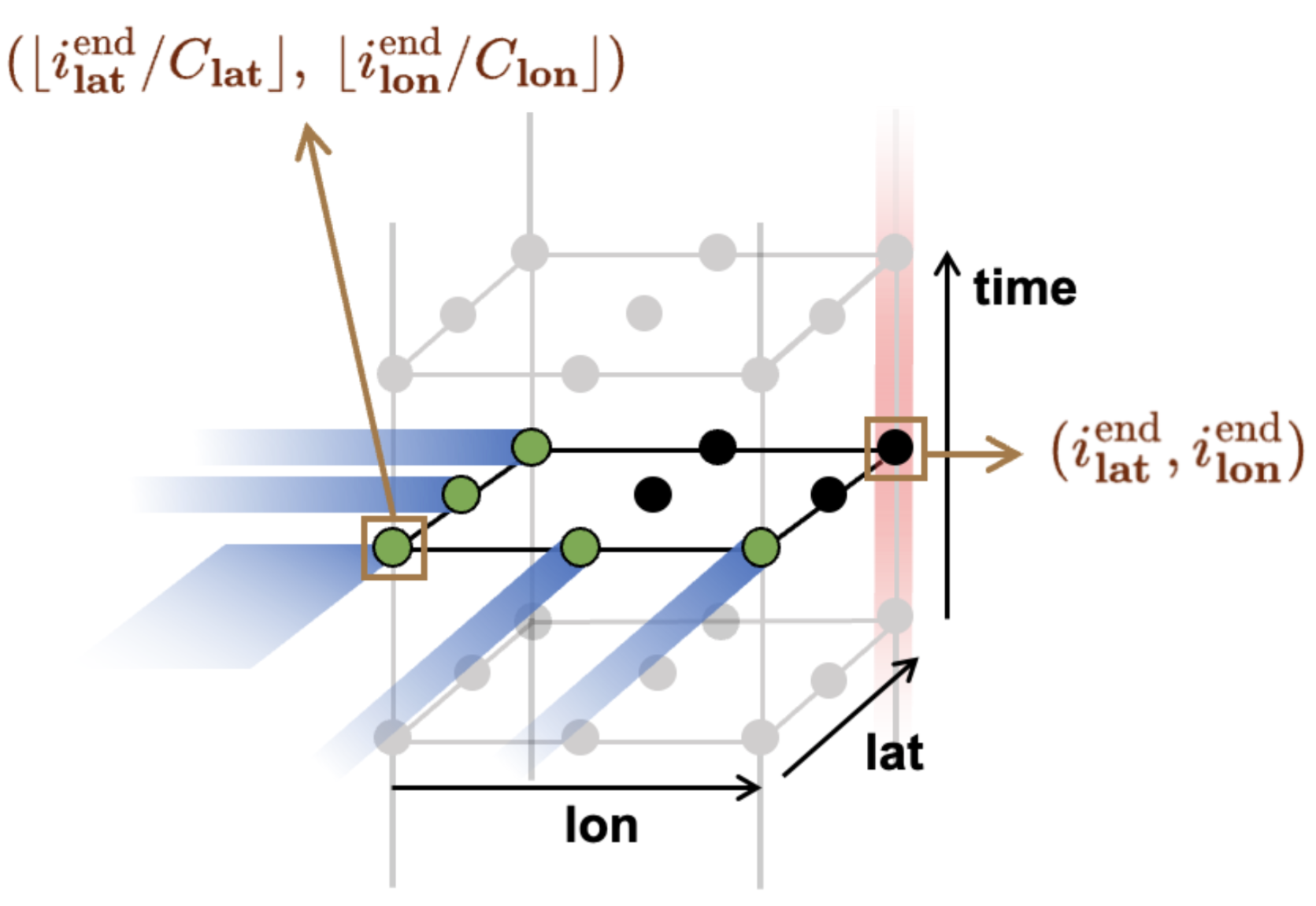}
\caption{Global area averaging computation using the ``composite'' chunk at a specific time slice. The average is obtained by summing the data values within the ``composite'' chunk at the endpoint – including 1-D accumulation data (green circles with blue gradient bars), 2-D accumulation data (green circle with blue gradient parallelogram), and raw data (black circles) -- and then divided by the total number of data points from the origin. Spatial indices referenced in Equation-4 are shown in brown.}
\label{fig_4}
\end{figure}

For more generalized area averaging when the averaging boundary is not aligned with the map boundary, accumulation-based averaging becomes more complex to represent; however, its computational complexity remains $O(1)$ along the averaging dimensions. In such cases, instead of processing a single chunk at the averaging endpoint, 2 or 4 chunks need to be processed at those corners not aligned with the map boundary. The aggregation result is still a linear combination of raw and accumulated values in these few corner chunks; however, unlike merely summing “composite” chunk values at the endpoints, subtraction is required at certain corners. The sign of the linear combination, either positive or negative, can be programmatically determined based on the chunk's position within the bounding box (Figure 5).

\begin{figure}[b]
\centering
\includegraphics[width=\linewidth]{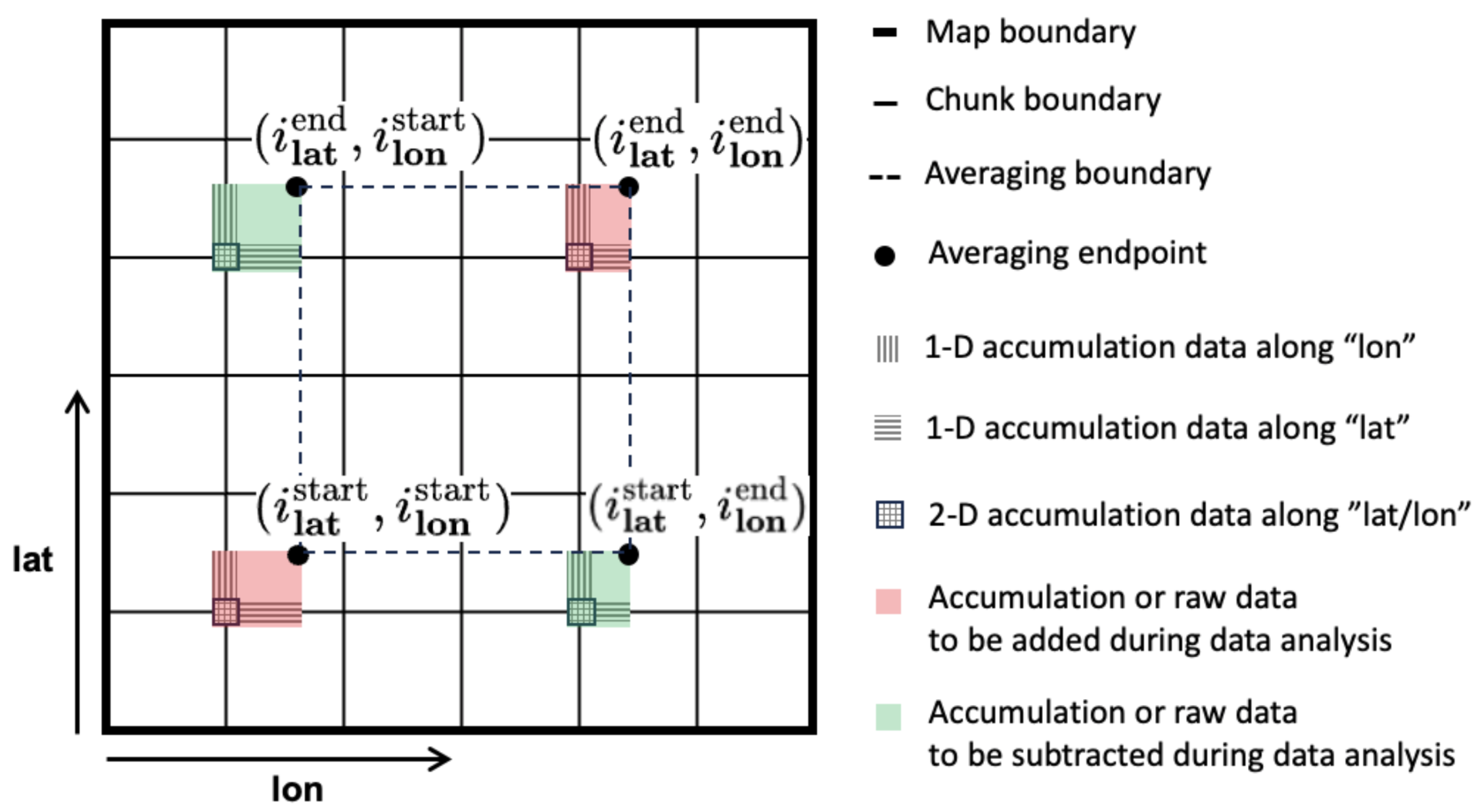}
\caption{General accumulation-based area averaging. When the averaging boundary is not aligned with the map boundary, ``composite'' chunks need to be processed at the non-aligned corners. As illustrated here and Equation-5, the averaging result is derived by subtracting the accumulation values at  $(i^{start}_{lat}, i^{end}_{lon})$ and $(i^{end}_{lat}, i^{start}_{lon})$  from that at $(i^{end}_{lat}, i^{end}_{lon})$, and then adding back that at $(i^{start}_{lat}, i^{start}_{lon})$  to compensate for the over-subtraction.}
\label{fig_5}
\end{figure}

\begin{align}
&\mathrm{avg}\big[(i_{\textbf{lat}}^{\text{start}}, i_{\textbf{lon}}^{\text{start}}), 
(i_{\textbf{lat}}^{\text{end}}, i_{\textbf{lon}}^{\text{end}})\big] = \notag \\
&\frac{
    \mathrm{sum}[i_{\textbf{lat}}^{\text{end}}, i_{\textbf{lon}}^{\text{end}}]
    - \mathrm{sum}[i_{\textbf{lat}}^{\text{start}}, i_{\textbf{lon}}^{\text{end}}]
    - \mathrm{sum}[i_{\textbf{lat}}^{\text{end}}, i_{\textbf{lon}}^{\text{start}}]
    + \mathrm{sum}[i_{\textbf{lat}}^{\text{start}}, i_{\textbf{lon}}^{\text{start}}]
}{
    (i_{\textbf{lat}}^{\text{end}} - i_{\textbf{lat}}^{\text{start}})
    \cdot (i_{\textbf{lon}}^{\text{end}} - i_{\textbf{lon}}^{\text{start}})
} 
\end{align}

\subsubsection{Time averaging}
The accumulation-based time-averaging of 3-D geospatial data is akin to area averaging, following the same fundamental workflow outlined in Figure 2. Figure 6 depicts a representative starting and ending chunk on the time-averaging boundary along the averaging dimension, with the averaging endpoints marked by the red planes. Since the averaging dimension in this case is {“time”}, the only non-empty dimension subset is {“time”} itself. Therefore, to construct the ``composite'' chunks, only those data points in the bottom plane need to be replaced by the corresponding 1-D accumulation values along the “time” dimension. The averaging result for each spatial coordinate remains a linear combination of the values within these ``composite'' chunks.

\begin{align}
\mathrm{avg}[i_{\textbf{time}}^{\text{start}}, i_{\textbf{time}}^{\text{end}}] &=  \\
&\Bigg( \mathrm{acc}_{\lfloor i_{\boldsymbol{\mathrm{time}}}^{\text{end}} / C_{\boldsymbol{\mathrm{time}}} \rfloor - 1}^{\boldsymbol{\mathrm{time}}} + \sum_{i = \lfloor i_{\boldsymbol{\mathrm{time}}}^{\text{end}} / C_{\boldsymbol{\mathrm{time}}} \rfloor}^{i_{\textbf{time}}^{\text{end}}} \mathrm{data}[i] \notag \\
\quad &- \mathrm{acc}_{\lfloor i_{\boldsymbol{\mathrm{time}}}^{\text{start}} / C_{\boldsymbol{\mathrm{time}}} \rfloor - 1}^{\boldsymbol{\mathrm{time}}} - \sum_{i = \lfloor i_{\boldsymbol{\mathrm{time}}}^{\text{start}} / C_{\boldsymbol{\mathrm{time}}} \rfloor}^{i_{\boldsymbol{\mathrm{time}}}^{\text{start}}} \mathrm{data}[i] \Bigg) \notag \\
\quad &\times \frac{1}{i_{\textbf{time}}^{\text{end}} - i_{\textbf{time}}^{\text{start}}}.
\end{align}


where $i^{start}_{time}$ and $i^{end}_{time}$ are the time indices of the starting and ending points, and $C_{time}$ is the chunk size along time dimension; it’s obvious that  $(\lfloor i_{\boldsymbol{\mathrm{time}}}^{\text{start}}/C_{\boldsymbol{\mathrm{time}}} \rfloor$ and $(\lfloor i_{\boldsymbol{\mathrm{time}}}^{\text{end}}/C_{\boldsymbol{\mathrm{time}}} \rfloor$ are the chunk indices of the starting and ending points along time dimension (see Figure 6). The latitude and longitude indices are omitted for clarity.

\begin{figure}[]
\centering
\includegraphics[width=\linewidth]{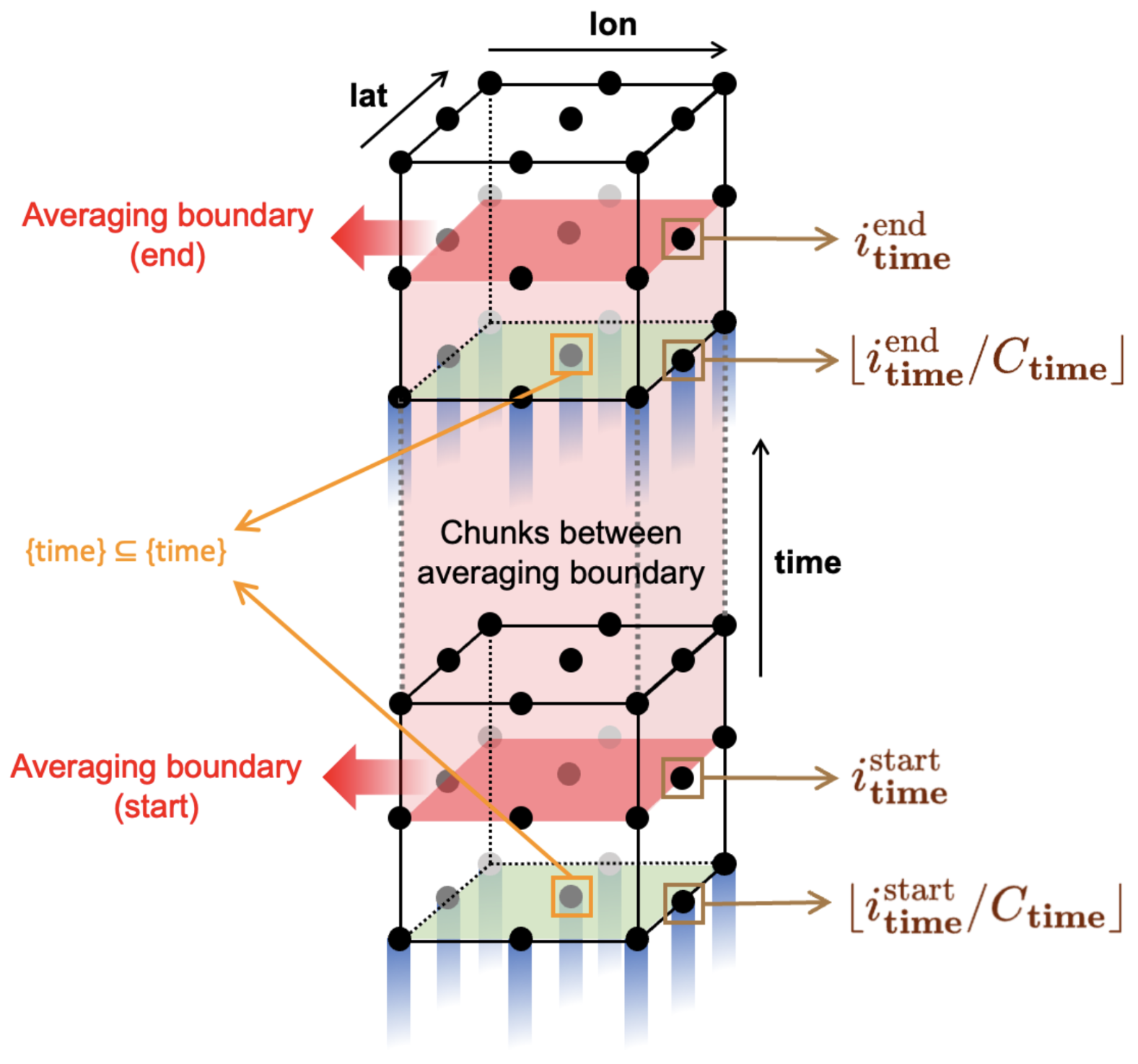}
\caption{General accumulation-based time averaging. This figure shows the two chunks encompassing the initial and final averaging boundaries (shown as the red planes), whereas the chunks between the averaging boundary are omitted (dashed lines). As shown in orange, the only non-empty subset of the averaging dimension set \{“time”\} is itself, and therefore only the 1-D accumulation from the foremost bottom plane (shown in green) are incorporated in the “composite” chunk construction. Time indices referenced in Equation-7 are shown in brown.}
\label{fig_6}
\end{figure}

\subsection{Zarr extension proposal}

The Zarr-based chunk-level accumulation method has two remarkable features:

\begin{enumerate}
    \item Generality. While the previous method description focuses on geospatial data, it’s noteworthy that this method is a generic approach. It is dimension-agnostic and can be applied to data with any type or number of dimensions.
    
    \item Tunability. Accumulation does not have to be calculated for each single chunk; instead, it can occur after multiple chunks (Figure 7). This is particularly beneficial for Zarr datasets with small chunks, because it can substantially reduce the total accumulation data size. However, it should be noted that when the aggregation boundary does not align with the accumulation chunk, multiple chunks adjacent to the endpoint need to be retrieved for data analysis, potentially impacting performance (Figure 7). Therefore, a trade-off exists between the accumulation interval and the performance: a larger accumulation interval reduces accumulation data volume but may also impact performance. Despite this, this method is still expected to significantly outperform brute-force aggregation in both efficiency and performance.
\end{enumerate}

These features facilitate the straightforward implementation of this method in general-purpose software applications. Accumulation specifications, such as the accumulation dimension names and intervals, can be stored as metadata or attributes alongside the accumulation data. These specifications can then be read by the program for accumulation data generation and analysis. This concept is analogous to how Xarray manages dimension names in multi-dimensional arrays (https://docs.xarray.dev/en/latest/getting-started-guide/why-xarray.html).

\begin{figure}[!t]
\centering
\includegraphics[width=\linewidth]{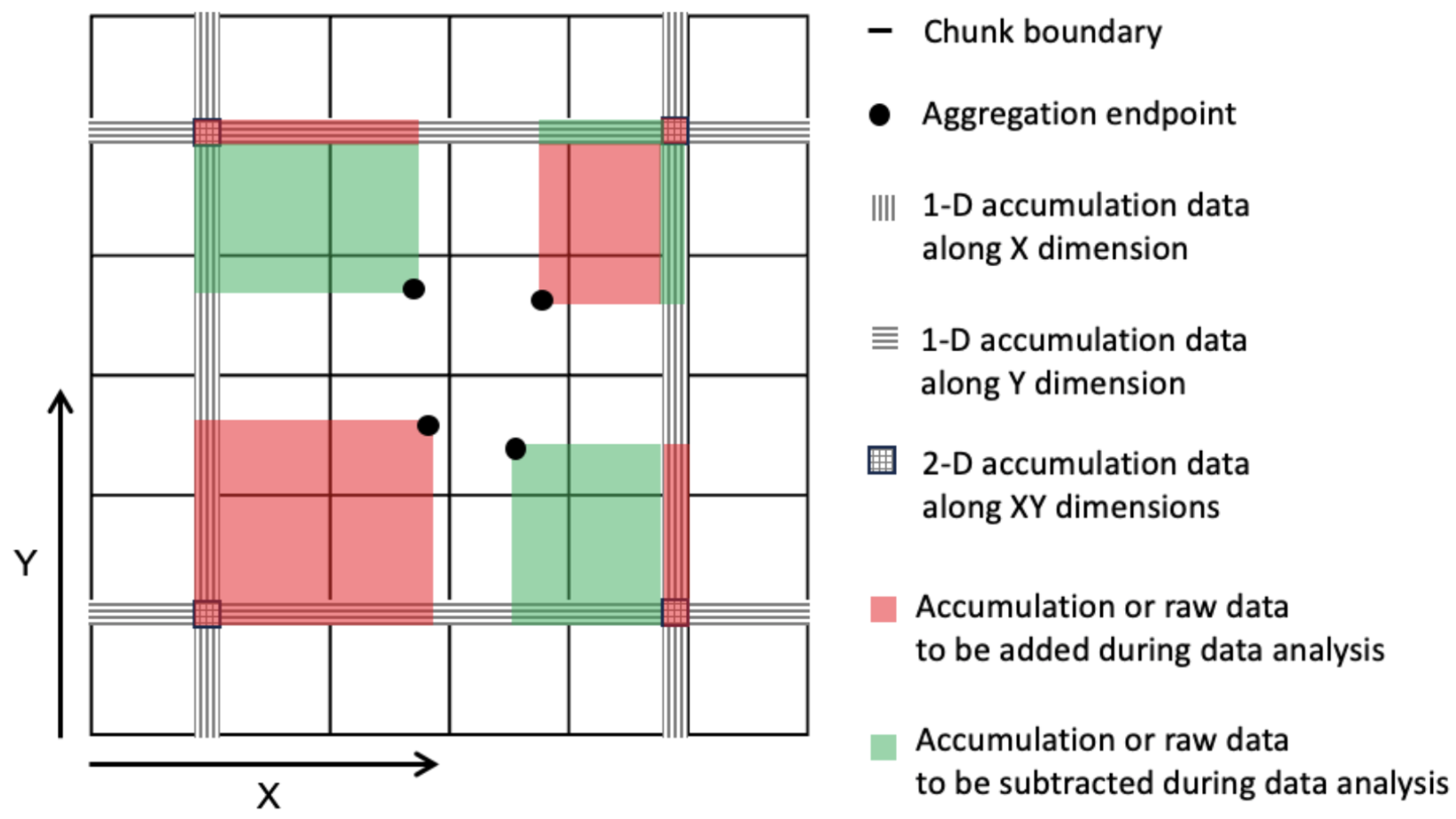}
\caption{Accumulation tunability. This figure illustrates a simple 2-D data where accumulation occurs after every 4 chunks along each dimension. When the aggregation endpoint does not align with the accumulation chunk, multiple chunks need to be retrieved for data analysis (in red or green). Depending on the specific aggregation endpoint location, the accumulation or raw data need to be either added (in red) or subtracted (in green) to derive the final results. The choice of addition or subtraction can be programmatically determined by the endpoint’s location, a process not detailed in this manuscript.}
\label{fig_7}
\end{figure}

In an effort to formalize and standardize this process, we recently proposed a Zarr extension defining the schema of the accumulation specifications. The extension proposal is available in the ``zarr-developers/zeps'' Github repo as ZEP0005 (https://github.com/zarr-developers/zeps/blob/main/draft/ZEP0005.md). The proposal details, including the structure of the accumulation data group and the schema for the accumulation data group and data sets, are summarized as follows:

\subsubsection{Zarr group structure of accumulation data}
This Zarr extension proposal assumes compliance of the raw Zarr data with the Xarray specification including the “\_ARRAY\_DIMENSIONS” attribute and dimension arrays. The accumulation datasets are organized in a data group adjacent to the raw data and dimension arrays with the following structure:

\begin{scriptsize}
\begin{verbatim}
+-- ${dimension_array}
+-- ...
+-- ${raw_dataset}
+-- ...
+-- ${raw_dataset}_accumulation_group
    +-- .zgroup
    +-- .zattr
    +-- ${accumulation_dataset_1}
    |   +-- .zarray
    |   +-- .zattr
    |   +-- ...
    +-- ${accumulation_dataset_2}
    |   +-- .zarray
    |   +-- .zattr
    |   +-- ...
    ...
\end{verbatim}
\end{scriptsize}

where ``\${dimension\_array}'' is the data array for the dimension variable, ``\${raw\_dataset}'' is the data array for the raw dataset, ``\${raw\_dataset\_accumulation\_group}'' is the group for accumulation, and ``\${accumulation\_dataset\_1}'' and ``\${accumulation\_dataset\_2}'' are the data arrays for each accumulation dataset.

\subsubsection{Zarr attribute file of accumulation group}
The accumulation group attribute file, ``\${raw\_dataset\_accumulation\_group/.zattr'', provides details of the accumulation implementation and data organization. The relevant portion of the schema for this attribute file is shown as follows:

\begin{scriptsize}
\begin{verbatim}
{
  "$schema": "http://json-schema.org/draft-07/schema#",
  "type": "object",
  "definitions": {
    "accumulation_data_array": {
      "type": "object",
      "properties": {
        "_DATA_UNWEIGHTED": {
          "type": "string"
        },
        "_DATA_WEIGHTED": {
          "type": "string"
        },
        "_WEIGHTS": {
          "type": "string"
        }
      },
      "patternProperties": {
        "^(?!_DATA_UNWEIGHTED|_DATA_WEIGHTED|_WEIGHTS).*$": {
          "$ref": "#/definitions/accumulation_data_array"
        }
      },
      "additionalProperties": false
    }
  },
  "properties": {
    "_ACCUMULATION_GROUP": {
      "type": "object",
      "patternProperties": {
        "^(?!_DATA_UNWEIGHTED|_DATA_WEIGHTED|_WEIGHTS).*$": {
          "$ref": "#/definitions/accumulation_data_array"
        }
      },
      "additionalProperties": false
    }
  },
  "required": [
    "_ACCUMULATION_GROUP"
  ]
}
\end{verbatim}
\end{scriptsize}

This schema includes a recursive definition located under the root key ``\_ACCUMULATION\_GROUP'' (referred to as ``\#/definitions/accumulation\_data\_array''), which provides details of the accumulation data, including the dataset names, accumulation data types, and dimensions. The keys within its properties represent the accumulation data types: ``\_DATA\_UNWEIGHTED'' and ``\_DATA\_WEIGHTED'' correspond to unweighted and weighted data accumulations, whereas ``\_WEIGHTS'' corresponds to the accumulation on the weights themselves. The values of its properties give the accumulation dataset names. Accumulation dimension names are stored within the ``patternProperties'' keys; it is noteworthy that these dimension names need to be ordered to avoid ambiguity and redundancy.

An example of the above Zarr attribute file is given as follows. This data has three dimensions including ``latitude'', ``longitude'', and ``time''. The accumulation is computed for the weighted data (``\_DATA\_WEIGHTED'') and weights (``\_WEIGHTS''). If we want to provide the time-averaged map and area-averaged time series, the accumulation is only needed for the dimension combinations of ``latitude'', ``longitude'', ``time'', and ``latitude+longitude''; all other dimension combinations (e.g., ``latitude+time'', ``longitude+time'', and ``latitude+longitude+time'') are therefore empty (``\{\}'').

\begin{scriptsize}
\begin{verbatim}
{
  "_ACCUMULATION_GROUP": {
    "latitude": {
      "_DATA_WEIGHTED": "acc_lat",
      "_WEIGHTS": "acc_wt_lat",
      "longitude": {
        "_DATA_WEIGHTED": "acc_lat_lon",
        "_WEIGHTS": "acc_wt_lat_lon",
        "time": {}
      },
      "time": {}
    },
    "longitude": {
      "_DATA_WEIGHTED": "acc_lon",
      "_WEIGHTS": "acc_wt_lon",
      "time": {}
    },
    "time": {
      "_DATA_WEIGHTED": "acc_time",
      "_WEIGHTS": "acc_wt_time"
    }
  }
}
\end{verbatim}
\end{scriptsize}

\subsubsection{Zarr attribute file of accumulation data array}
As highlighted earlier, the feature referred to as ``tunability'' allows for accumulation to be performed after every certain number of chunks (instead of for each single chunk) to further reduce the accumulation data size. In this proposal, this tunable parameter is termed the ``accumulation stride''. This parameter is stored under the ``\_ACCUMULATION\_STRIDE'' property in the Zarr attribute file associated with the accumulation dataset (e.g., ``\${raw\_dataset}\_accumulation\_group/\{accumulation\_dataset\_1\} \newline /.zattr''), along with the dimension labels following the Xarray convention (``\_ARRAY\_DIMENSIONS''). The relevant portion of the schema for this attribute file is shown as follows:

\begin{scriptsize}
\begin{verbatim}
{
  "$schema": "http://json-schema.org/draft-07/schema#",
  "type": "object",
  "properties": {
    "_ARRAY_DIMENSIONS": {
      "type": "array",
      "items": {
        "type": "string"
      }
    },
    "_ACCUMULATION_STRIDE": {
      "type": "array",
      "items": {
        "type": "integer"
      }
    }
  },
  "required": [
    "_ARRAY_DIMENSIONS",
    "_ACCUMULATION_STRIDE"
  ]
}
\end{verbatim}
\end{scriptsize}

The ``\_ARRAY\_DIMENSIONS'' and ``\_ACCUMULATION\_STRIDE'' arrays must be of equal length. Each element in the ``\_ACCUMULATION\_STRIDE'' array corresponds to the accumulation stride along the respective dimension specified at the same index in the ``\_ARRAY\_DIMENSIONS'' array. The value of accumulation stride should be a non-negative integer: a positive value represents the specified accumulation stride, whereas a value of 0 indicates the accumulation is not performed along the given dimension.

For example, the following attribute file represents the accumulation that is performed along only the time dimension every other chunk:

\begin{scriptsize}
\begin{verbatim}
{
  "_ARRAY_DIMENSIONS": [
    "latitude",
    "longitude",
    "time"
  ],
  "_ACCUMULATION_STRIDE": [
    0,
    0,
    2
  ]
}
\end{verbatim}
\end{scriptsize}

and the following attribute file represents the accumulation that is performed along the ``latitude'' dimension for each chunk, and along ``longitude'' dimension every 3 chunks:

\begin{scriptsize}
\begin{verbatim}
{
  "_ARRAY_DIMENSIONS": [
    "latitude",
    "longitude",
    "time"
  ],
  "_ACCUMULATION_STRIDE": [
    1,
    3,
    0
  ]
}
\end{verbatim}
\end{scriptsize}

\subsubsection{Application Interface}
The accumulation-based workflow (Figure 2) requires the application to locate the accumulation data along certain dimensions. The accumulation data array name for the given dimensions can be obtained from the accumulation group attributes. The following example in Python shows the steps to get the weighted accumulation data array name along “latitude” + “longitude” dimensions:

\begin{figure}[!h]
\centering
\includegraphics[width=\linewidth]{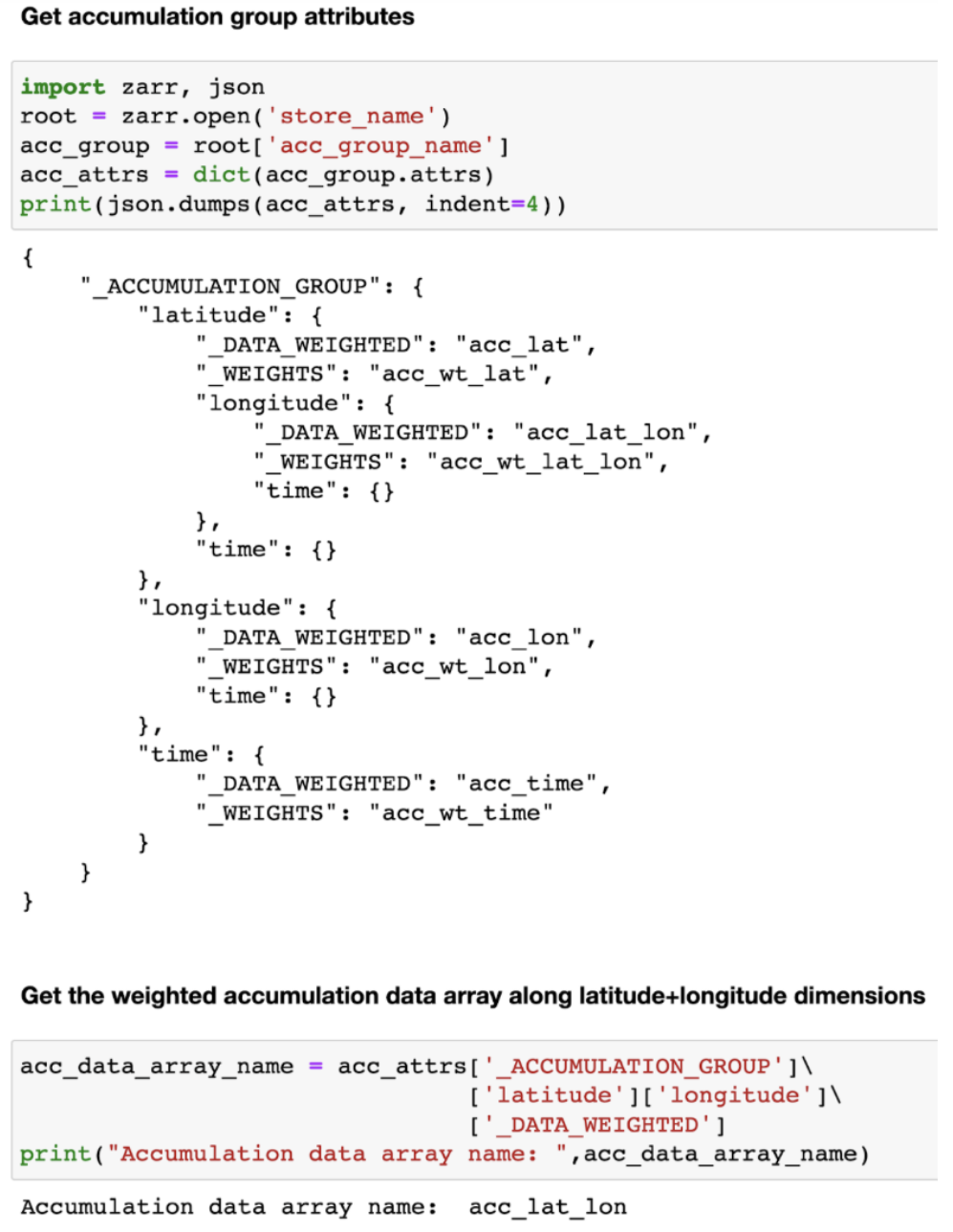}
\label{interface_1}
\end{figure}

The accumulation stride is also needed to locate the accumulation data for a given chunk number. They can be obtained from the accumulation data attributes, and the following example shows the steps to get the accumulation stride for the accumulation data along latitude+longitude dimensions:

\begin{figure}[h]
\centering
\includegraphics[width=\linewidth]{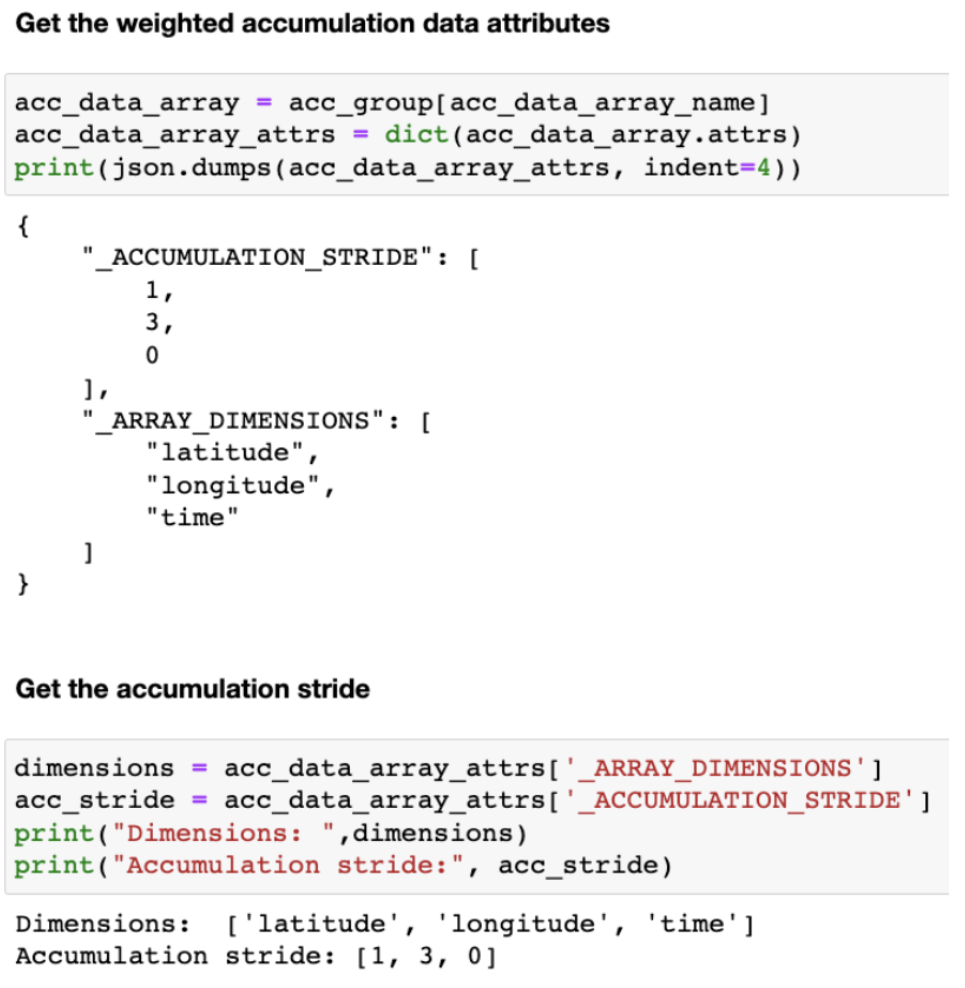}
\label{interface_2}
\end{figure}

\section{Results}
Based on the chunk-level accumulation method described above, we generated accumulation data from a high-resolution Earth data product. Using the accumulation data, we then performed large region time- and area-averaging computations, and compared the performance of this method against the brute-force approach. Our benchmark was the ‘precipitationCal’ variable from the GPM IMERG Final Precipitation L3 (GPM\_3IMERGHH) Version 6 product, which provides global surface precipitation estimates at a spatial resolution of 0.1 degrees and a temporal resolution of 0.5 hours. Please note that this product has recently been superseded by Version 07 \cite{huffman_2023}.

\subsection{Accumulation Data Generation}
The Zarr store for the raw GPM\_3IMERGHH data is organized with chunk sizes of 36 and 72 along the latitude and longitude dimensions, respectively, resulting in 50 chunks per dimension, and a chunk size of 200 along the time dimension. The accumulation data to be generated includes 1-D accumulation along the time, latitude, and longitude dimensions, as well as 2-D accumulation across the latitude/longitude plane, as required by time- and area-averaging services. To optimize storage, accumulation data was generated every 2 chunks along the latitude, longitude, and time dimensions rather than for each individual chunk, resulting in a smaller accumulation data size with minimal impact on computational performance (Figure 7). The script for generating accumulation data, which adheres to the Zarr extension proposal described above, is available in the GitHub repository: \url{https://github.com/nasa/zarr-accumulation-generation}.

Accumulation data generation was conducted on an AWS EC2 instance of type r5.16xlarge (64 vCPUs, 512 GiB memory) at an on-demand rate of \$4.032 per hour. The entire process took 1 hour to generate accumulation data for 20 years (\~350,000 time slices), resulting in an accumulation dataset approximately 5\% the size of the raw data.

\begin{figure}[!t]
\centering
\includegraphics[width=\linewidth]{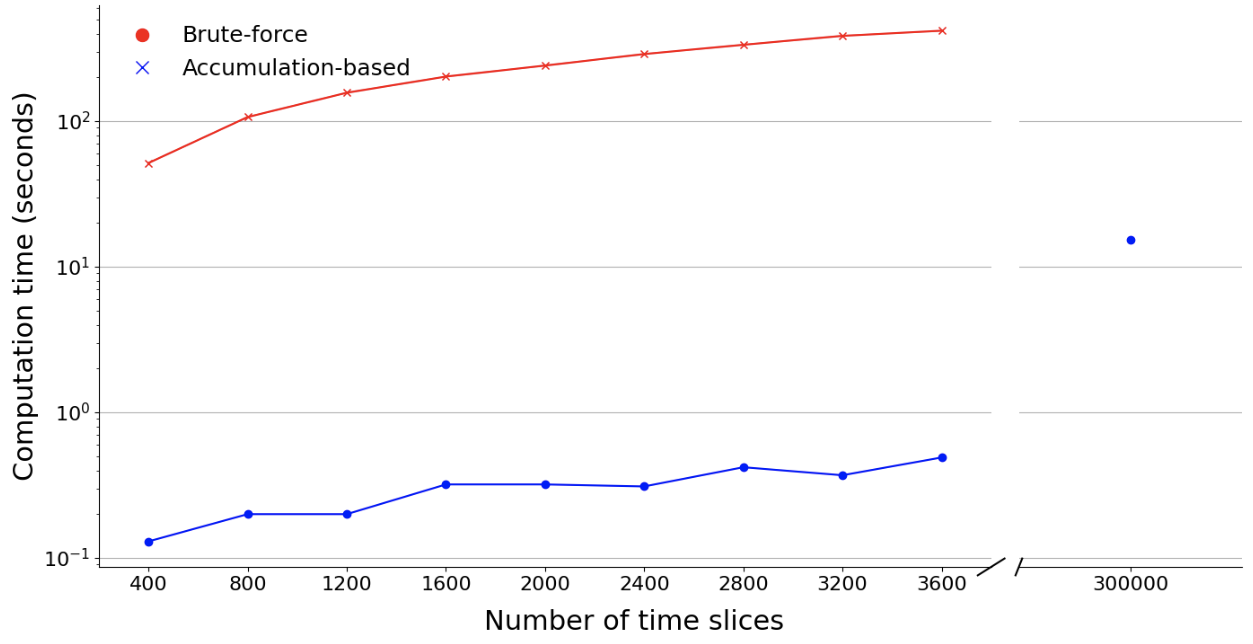}
\caption{Performance comparison of the global-averaged time series computation between the accumulation-based (blue) and brute-force (red) methods. The computations were performed using the `precipitationCal' variable from the GPM\_3IMERGHH Version 6 product as a benchmark. All computations were conducted on an AWS EC2 instance with Dask version 2021.08.0 and Zarr version 2.16.1. Due to the extremely long processing time on this EC2 instance, the brute-force method was not attempted for 300,000 time slices.}
\label{fig_8}
\end{figure}

\begin{figure}[!t]
\centering
\includegraphics[width=\linewidth]{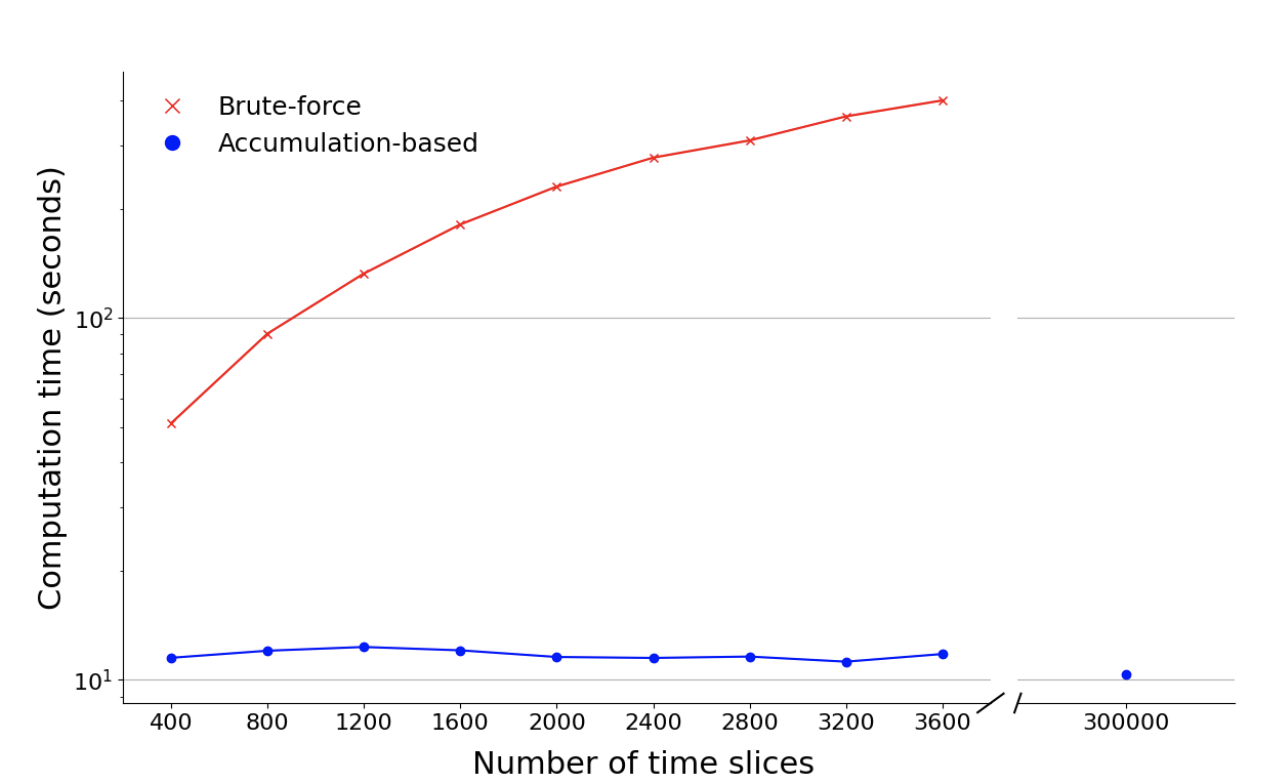}
\caption{Performance comparison of the global time-averaged map computation between the accumulation-based (blue) and brute-force (red) methods. The computations were performed using the ‘precipitationCal’ variable from the GPM\_3IMERGHH Version 6 product as a benchmark. All computations were conducted on an AWS EC2 instance with Dask version 2021.08.0 and Zarr version 2.16.1. Due to the extremely long processing time on this EC2 instance, the brute-force method was not attempted for 300,000 time slices.}
\label{fig_9}
\end{figure}

\subsection{Accumulation Data Analysis}
Based on the accumulated data prepared above, we performed area-averaged time series and time-averaged map computation, and compared their performance and accuracy against the brute-force method based solely on the raw data. The computations were conducted on an AWS EC2 r5.16xlarge instance (64 vCPUs, 512 GiB memory) with Dask version 2021.08.0 and Zarr version 2.16.1. The on-demand price for this EC2 instance type is \$4.032 per hour.

\subsubsection{Area-averaged time series}
The area-averaged time series was computed globally, covering 400 to 300,000 time slices. Figure 8 shows the performance comparison of the global-average time series computation between the accumulation-based and brute-force methods. For time slices ranging from 400 to 3,600, the accumulation-based method achieved a significant speedup of approximately 3 orders of magnitude. Notably, for the global-averaged time series of 300,000 time slices (equivalent to approximately two years of data), the accumulation-based method completed in 15 seconds, which is infeasible for the brute-force method on this EC2 instance.

\subsubsection{Time-averaged map}
The time-averaged map was computed by averaging over 400 to 300,000 time slices across the entire globe. Figure 9 shows a performance comparison of the time-averaged map computation between the accumulation-based and brute-force methods. Similar to the area-averaged time series, the accumulation-based method significantly outperforms the brute-force method. Specifically, from 400 to 3,600 time slices, the computation time for the brute-force method scaled linearly with the number of time slices (increasing from 50 to 400 seconds); however, the computation time for the accumulation-based method remained constant at approximately 10 seconds. This constant time was maintained even for 300,000 time slices (equivalent to approximately two years of data), which is infeasible for the brute-force method on this EC2 instance.

\subsubsection{Accuracy}
Based on the benchmark described above, Table 1 provides a comparison of the normalized RMSD (NRMSD) between the accumulation-based and brute force methods for the area-averaged time series and the time-averaged map. The NRMSD value is computed using the following formula:
\begin{equation}
NRMSD = \frac{\sqrt{\frac{1}{N}\sum_{i=1}^{N}(average_i^{brute} - average_i^{acc})^2}}{max(average^{brute})-min(average^{brute})}
\end{equation}
Where $average^{brute}_{i}$ is the average value computed using the brute force method at index $i$, $average^{acc}_{i}$ is the average value computed using the accumulation-based method at index $i$, $N$ is the total number of data points, and $max(average^{brute})$ and $min(average^{brute})$ represent the maximum and minimum average values computed using the brute force method.

\begin{table*}[]
\begin{tabular}{ccrrrrrrrrr}
\hline
\multicolumn{2}{|c|}{\textbf{Number of time slices}}                                                            & \multicolumn{1}{r|}{400}     & \multicolumn{1}{r|}{800}     & \multicolumn{1}{r|}{1200}    & \multicolumn{1}{r|}{1600}    & \multicolumn{1}{r|}{2000}    & \multicolumn{1}{r|}{2400}    & \multicolumn{1}{r|}{2800}     & \multicolumn{1}{r|}{3200}    & \multicolumn{1}{r|}{3600}    \\ \hline
\multicolumn{1}{|c|}{\multirow{2}{*}{\textbf{NRMSD}}} & \multicolumn{1}{c|}{\textbf{Area-averaged time series}} & \multicolumn{1}{r|}{1.15E-7} & \multicolumn{1}{r|}{1.17E-7} & \multicolumn{1}{r|}{1.14E-7} & \multicolumn{1}{r|}{1.15E-7} & \multicolumn{1}{r|}{1.14E-7} & \multicolumn{1}{r|}{1.15E-7} & \multicolumn{1}{r|}{1.01E-7}  & \multicolumn{1}{r|}{1.03E-7} & \multicolumn{1}{r|}{1.03E-7} \\ \cline{2-11} 
\multicolumn{1}{|c|}{}                                & \multicolumn{1}{c|}{\textbf{Time-averaged map}}         & \multicolumn{1}{r|}{1.44E-9} & \multicolumn{1}{r|}{3.25E-9} & \multicolumn{1}{r|}{1.62E-9} & \multicolumn{1}{r|}{1.59E-9} & \multicolumn{1}{r|}{1.58E-9} & \multicolumn{1}{r|}{1.51E-9} & \multicolumn{1}{r|}{6.32E-10} & \multicolumn{1}{r|}{1.89E-9} & \multicolumn{1}{r|}{1.94E-9} \\ \hline
\multicolumn{1}{l}{}                                  & \multicolumn{1}{l}{}                                    & \multicolumn{1}{l}{}         & \multicolumn{1}{l}{}         & \multicolumn{1}{l}{}         & \multicolumn{1}{l}{}         & \multicolumn{1}{l}{}         & \multicolumn{1}{l}{}         & \multicolumn{1}{l}{}          & \multicolumn{1}{l}{}         & \multicolumn{1}{l}{}        
\end{tabular}
\caption{Normalized root mean square deviation between the accumulation-based and brute force methods}
\label{table_results}
\end{table*}

The results demonstrate that the accumulation-based averaging computation agrees closely with the brute force method.

\section{Conclusion}
In this paper, we introduced and detailed a generic method to perform fast and cost-efficient data analysis on massive multi-dimensional data, such as high-resolution large-region time averaging or area averaging for geospatial data. This method is general-purpose, but particularly well-suited for data stored in chunked, cloud-optimized formats, such as Zarr, and for services running in distributed or cloud environments. It generates a minimal and size-tunable supplementary dataset that stores the cumulative sums along specific subset dimensions, which facilitates quick and cheap high-resolution, large-region data analysis, making it feasible with small instances or even microservices in the cloud.

This method has two notable features: the generality, being applicable to any type of data or number of dimensions, and the tunability, allowing for a balance between accumulation data size and performance. We proposed a Zarr extension that formalizes the schema for accumulation specifications, aiming to standardize and simplify the implementation of this method in general-purpose software applications. 

We performed the benchmark tests of this method on a high resolution geospatial dataset, and compared the data analysis performance of this accumulation method against a brute-force approach. Our method, which added just 5\% more data for accumulation, significantly outperformed the brute-force method by approximately 3 orders of magnitudes while maintaining accuracy. The cost for accumulation-based service is also minimal for both accumulation data generation and analysis, making it potentially feasible for cloud microservices such as AWS lambda. 

In summary, this approach represents a cost-effective and scalable solution for large-scale data analysis while preserving the accuracy of the brute-force method. This approach is also suited for pre-computing statistics such as min, max, and sum of squares, which can facilitate more comprehensive data analysis and statistical computations. With further refinement, this method has the potential for broader application in real-time data analysis and other domains involving multi-dimensional datasets, supporting a wider range of scalable scientific use cases beyond Earth science data analysis.

\newpage

\end{document}